# The Athena Data Dictionary and Description Language

Alain Bazan, Thierry Bouedo, Philippe Ghez
*LAPP, Annecy-le-Vieux, FRANCE*

Massimo Marino, Craig Tull
*LBNL, Berkeley, USA*

Athena is the ATLAS off-line software framework, based upon the GAUDI architecture from LHCb. As part of ATLAS' continuing efforts to enhance and customise the architecture to meet our needs, we have developed a data object description tool suite and service for Athena. The aim is to provide a set of tools to describe, manage, integrate and use the Event Data Model at a design level according to the concepts of the Athena framework (use of patterns, relationships, ...). Moreover, to ensure stability and reusability this must be fully independent from the implementation details. After an extensive investigation into the many options, we have developed a language grammar based upon a description language (IDL, ODL) to provide support for object integration in Athena. We have then developed a compiler front end based upon this language grammar, JavaCC, and a Java Reflection API-like interface. We have then used these tools to develop several compiler back ends which meet specific needs in ATLAS such as automatic generation of object converters, and data object scripting interfaces. We present here details of our work and experience to date on the Athena Definition Language and Athena Data Dictionary.

## 1. INTRODUCTION

This document describes in brief the development and implementation of an "ATLAS Data Dictionary" (ADD) in the Athena Architecture. For full details see:[1].
Athena is the ATLAS off-line software framework, based upon the GAUDI architecture from LHCb. As part of ATLAS' continuing efforts to enhance and customise the architecture to meet needs of the users, we have developed a data object description tool suite and service for Athena. The term "data dictionary" is being used in ATLAS to cover several related, but distinct concepts and techniques. Each of these concepts plays a different set of roles in an architecture dependent upon a data dictionary. We categorise these concepts into three general topics:

- **Introspection/Reflection/Object Description/Run-Time Typing:** This refers to objects in program memory with the ability to describe themselves in a programmatic way through a public API such that they can be manipulated without a priori knowledge of the specific class/type of the object.
- **Code Generation:** This refers to a process of generating code for performing a specific task from a generic description/input file.
- **Self-Describing External Data Representation (e.g. Data Files):** This refers to external data representations (e.g. file formats, on-wire data formats) which contain metadata describing the payload of the data file, etc...

## 2. ADVANTAGE OF A DATA DICTIONARY

The data dictionary is a description of the objects to a high abstraction level.

These tool avoid tedious integration of objects to the framework, concentrate the object development only on his behaviour and provide the objects with all the mechanism of conversion between transient and persistent stores. At run time it gives access to transient objects allowing debugging, visualisation, use scripting... These description allows re-use of the objects already present in the dictionary, the management of the evolution of the described objects and provides information on persistent objects and collections without loading them in transient store.

## 3. LANGUAGE AND TOOLS

### 3.1. Choice

One of the most visible implementation decisions of a DD for Athena is the choice of the computer language used in the dictionary. Declarative computer languages are widely used tools in the CS and IT communities. A list of

|  | C++ | IDL | JAVA | ODL | DDL | XML |
|---|---|---|---|---|---|---|
| Machine Independence | No | Yes | Yes | Yes | Yes | Yes |
| Programming Language Independence | No | Yes | No | Yes | No | Yes |
| Open-Source/Free Parsers Available | Yes | Yes | Yes | Yes | No | ? |
| Object Behavior Definable | Yes | Yes | Yes | Yes | Yes | ? |
| Object State Definable | Yes | Yes | Yes | Yes | Yes | ? |
| Public/Private Member | Yes | No | Yes | No | Yes | ? |
| Persistency | No | No | Yes | No | Yes | ? |
| Use of Predefined Types | Yes | Yes | Yes | Yes | Yes | ? |
| Use of External (Undefined) Types | No | No | No | No | No | ? |

Table 3: ADL candidate feature comparison

choices considered and associated tools available to parse the language is shown in the comparison's Table 3:

After an extensive investigation into the many options, we concluded that none of the language candidates fully matched ATLAS requirements, and that some compromise and/or language extension would be required.

We settled on and developed a language grammar based upon a proper subset of IDL 2.0 extended to provide support for object persistency and complex inter-object relationships.

The included extensions are:
ODL keyword to express bi-directional relationships: *relationship*
keyword to express persistency: *persistent*
keyword to support opaque objects: *extern*
keyword to declare objects of Athena: *DataObject, ContainedObject, CollectionObject* keyword to manage the visibility of the objects attributes: *private*

We called this extended proper subset of IDL: ADL for Athena Description Language. Such a declarative language helps separating objects' interfaces and behaviours from their implementations, isolating users of a system from implementation details, facilitating technology migration, and easing software development by eliminating tedious and error-prone rote programming.

Moreover, the choice of ADL because of its explicit independence of programming languages makes future possible evolution more feasible.

### 3.2. Tools

Code generation tools are parser-based tools which process the ADL. With the choice of a real computer language as the basis of the Data Dictionary, it becomes imperative that a real parser be used to compile the DD language and realise the DD functionality. Experience has shown that multiple back-ends (emitters) for the parser are necessary. The reality of a possible evolution of ADL suggests that the compiler front-end should be replaceable.

We chose JavaCC (the Java Compiler Compiler) as the parser for our compiler front end for the following reasons:
- Large number of languages supported (34 grammars from Ada to XML)
- Widely used & actively supported and developed
- Easily extended grammar
- Platform independence

Of all the tools considered and evaluated, JavaCC was the only one which supported all of the candidate languages. This made it particularly attractive in that a change in ADL language does not imply a change in parser.

## 4. DESIGN

### 4.1. From description to utilisation

The high-level design of the code generator is a standard 2-tier design. An ADL object description is fed into the Compiler Front End (CFE) consisting of the JavaCC generated parser (from the ADL Grammar). The parser produces an Abstract Syntax Tree using the JJTree package. A standard visitor pattern class walks the AST and fills an in-memory representation of the object description (the Meta-Object Representation). Multiple Compiler Back Ends (CBEs) use the information stored in the Meta-Object Representation to generate code for use in the Athena framework.

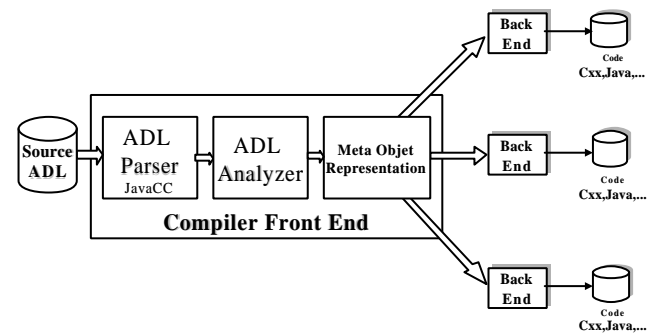

### 4.2. Meta-model

The Meta-Object Representation is a set of classes implementing a Java Reflection-like API and which insulates the writers of the compiler back-ends (CBEs) from the details of the JJTree AST. The static class diagram including the Meta-Object Representation design is shown in Figure 4.2.

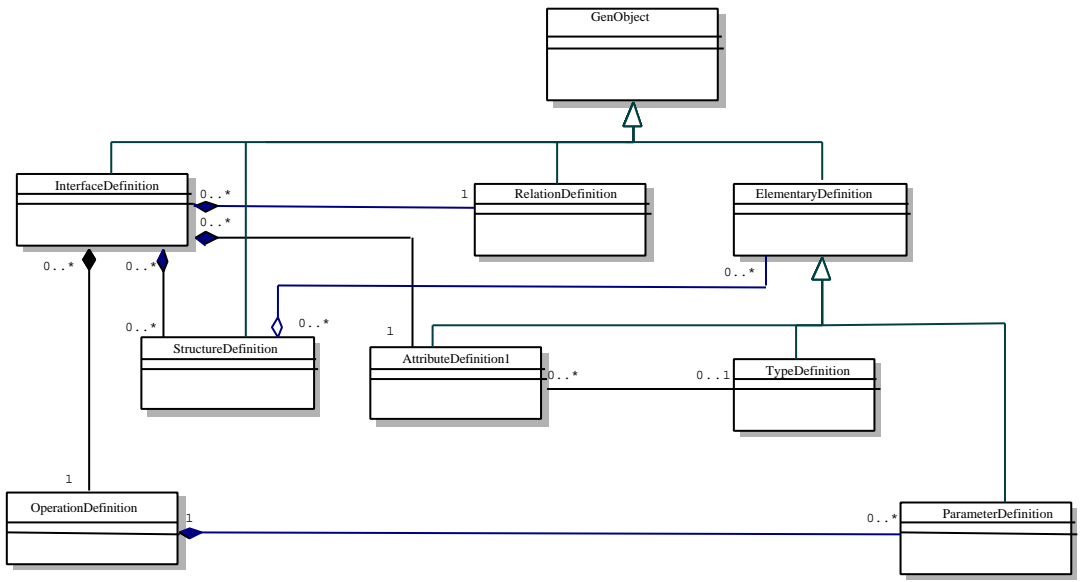

Figure 4.2 UML static class diagram

## 5. FUNCTIONALITIES

### 5.1. Code generation

The last code for the ATLAS data dictionary was released at the end of November 2002. Included in this release was the full ADL JavaCC grammar and generated parser, the JJTree-based visitor and Meta-Object Representation classes, and three compiler back ends. The following use cases diagram shows how to generate code:

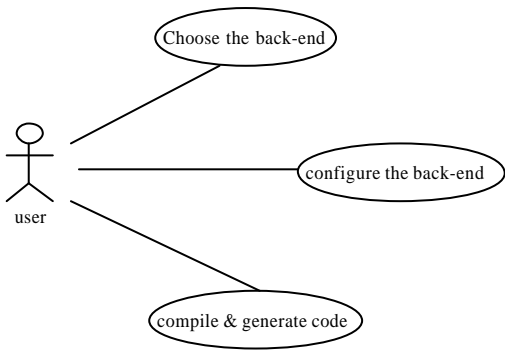

Although it is easy to write a new back-end as needed, the three following Compiler Back Ends are provided today:

• Data Object CBE: Generates C++ classes for user data objects with: ATLAS defined Constructors and Destructors, Single and Multiple Inheritance, Private Data Members & Accessor/Mutator Methods, Public Method Functions (beyond accessors/mutators), Interobject Associations, STL Support, and user written extensions.

• Converter CBE: Generates Athena converters for persistency using Objectivity Conversion service or ROOT conversion service.

• Scripting CBE: Provides a Python interface allowing limited access to, and control of data objects at the command line (see CHEP'01 paper 3-064). The three back ends work together or independently to provide needed Athena functionalities.

### 5.2. Dynamic interaction

Another main functionality of the Athena Data Dictionary is to dynamically manage the described objects. It answers the use cases as shown in the following diagram:

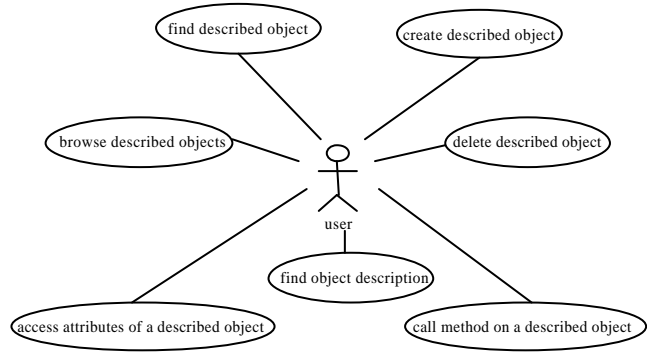

This functionality is mainly based on an Introspection/Reflection mechanism allowing connections between object's description and object's instance at run-time. This refers to objects in program memory with the ability to describe themselves in a programmatic way through a public API such that they can be manipulated without a priori knowledge of the specific class/type of the object.

This functionality should integrated and used in Athena according to the sequence shown in the figure 5.3.

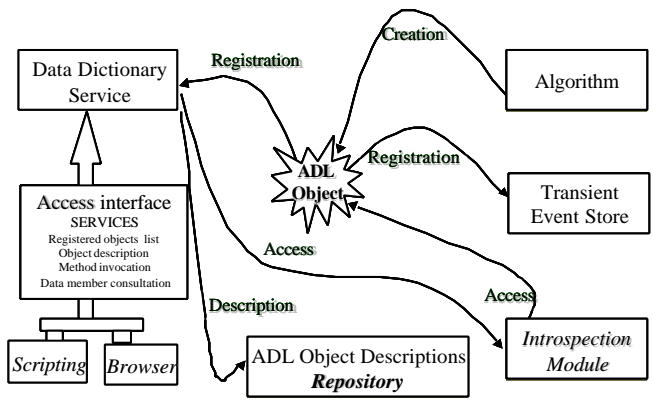

Figure 5.3 Dynamic interaction

The sequence goes through the following steps:
1. Creation of the object by the algorithm
2. Registration to the transient store and the data dictionary service
3. Access to the object description by the interactive part of the framework
4. Access to the object through the data dictionary service and the introspection module

## 6. CONCLUSION

The data dictionary-based code generators have been successfully used by some ATLAS collaborators, and three tutorials were given in June 2001, March 2002 and May 2002 based upon a Liquid Argon reconstruction data model. Integration in Athena has also been done by writing CMT fragments and Automatic ClassID generation. Connected to the ADL, a module has been developed for the Together case tool to generate graphically ADL code [2]. Moreover, a large amount of documentation (user guide, language reference manual, pocket guide, examples, FAQ,…) has been produce to provide user support.

Nevertheless, although this data dictionary project was answering the Atlas requirements, it has been abandon. This implies to ask ourselves about the reasons to draw lessons from that:
- Are people really ready to concentrate there efforts at the **design level** using an high level description language, independent of the implementation?
- Has this tool taken place too late in the Athena framework while a lot of C++ code was already written? (connected feedback: reverse engineering is not miraculous!).
- Has this project been politically killed at the birth of the LCG?

## References

[1] A. Bazan, T. Bouedo, P.Ghez, M.Marino, C.E.Tull, "Athena Web site - Dictionary", http://atlas.web.cern.ch/Atlas/GROUPS/ SOFTWARE/OO/architecture/DataDictionary/.

[2] M.Marino "Extending the code generation capabilities of the Together CASE tool to support Data Definition languages", 2003 Computing in High Energy and Nuclear Physics (CHEP03), La Jolla, CA, USA, 2003